\documentclass[aip,jcp,preprint]{revtex4-1}

\usepackage{bm}
\usepackage{amsmath,amssymb}

\providecommand{\eqnref}[1]{Eq.\ \eqref{#1}}
\renewcommand{\vec}[1]{\ensuremath{\mathbf{#1}}}

\providecommand{\ket}[1]{\ensuremath{\lvert#1\rangle}}
\providecommand{\bra}[1]{\ensuremath{\langle #1\rvert}}
\providecommand{\braket}[3]{\ensuremath{
 \langle #1\lvert #2\rvert #3\rangle}}
\providecommand{\mat}[1]{\ensuremath{\bm{\mathsf{#1}}}}

\providecommand{\bohr}{\ensuremath{\mu_\text{B}}}
\providecommand{\nbohr}{\ensuremath{\mu_\text{N}}}

\begin{document}
\title{Paramagnetic NMR chemical shift in a spin state subject to zero-field
splitting}
\author{Alessandro Soncini}
\email{asoncini@unimelb.edu.au}
\author{Willem Van den Heuvel}
\affiliation{School of Chemistry, The University of Melbourne, VIC 3010,
Australia}

\date{\today}

\begin{abstract}
We derive a general formula for the paramagnetic NMR nuclear
shielding tensor of an open-shell molecule in a pure spin state, subject 
to a zero-field splitting (ZFS). Our findings are in contradiction with a previous proposal.
We present a simple application of the newly derived formula to the case 
of a triplet ground state split by an easy-plane ZFS spin Hamiltonian. When $kT$ is much 
smaller than the ZFS gap, thus a single non-degenerate level is thermally populated, our 
approach correctly predicts a temperature-independent paramagnetic 
shift, while the previous theory leads to a Curie temperature dependence.
\end{abstract}

\maketitle

%\section{Introduction}

The nuclear magnetic resonance (NMR) chemical shift in molecular species
with an open-shell electronic structure is mainly governed by a term known as
the `paramagnetic shift', a temperature-dependent term arising from the
internal magnetic fields generated by the unpaired spin and unquenched orbital 
moments of the thermally populated Zeeman-split electronic degenerate ground state.
Despite the increasingly central role played by paramagnetic NMR in the elucidation 
of the structure of metallo-proteins\cite{Bertini}, and in the investigation of the spin dynamics 
in novel magnetic materials\cite{CarrettaBook2007}, only quite recently rigorous theories have 
been developed for the ab initio calculation of the paramagnetic 
NMR chemical shift\cite{Moon2004,Pennanen2008,VandenHeuvel2012,VandenHeuvel2012a}.

Of particular relevance in this respect is the work of Moon and Patchkovskii\cite{Moon2004},
that of Pennanen and Vaara\cite{Pennanen2008}, and that of Van den Heuvel and Soncini\cite{VandenHeuvel2012,VandenHeuvel2012a}.  
Moon and Patchkovskii derived an expression for the paramagnetic shielding tensor 
of a spin doublet state in terms of its $g$- and $A$-tensors \cite{Moon2004}. This treatment 
was extended by Pennanen and Vaara to arbitrary spin states, in the limit of 
weak spin-orbit coupling\cite{Pennanen2008}, and later generalised by us to a theory
that is valid for arbitrary strength of spin-orbit coupling, and arbitrary size 
of the degenerate manifold\cite{VandenHeuvel2012,VandenHeuvel2012a}. 
In Ref. \ \onlinecite{Pennanen2008} the authors also proposed a general formula for the paramagnetic 
shielding tensor of a spin state subject to zero-field splitting (see Eq. (10) in Ref.\ \onlinecite{Pennanen2008}).  
In this communication we present an alternative derivation of this formula, based 
on the general theory of NMR chemical shift we have recently developed\cite{VandenHeuvel2012,VandenHeuvel2012a}. 
Interestingly, we obtain a result that differs from that proposed in Ref.\ \onlinecite{Pennanen2008}.
In the last part of this communication we point out the difference, and argue for the correctness 
of our proposal by way of a simple example.

%\section{General theory}

We consider a molecule in the frozen nuclei approximation (also known as the
`solid state limit' of NMR \cite{McConnell1958,Kurland1970}). Assuming that the
zero-field splitting in the degenerate ground state is much smaller than the
energy separating the ground state from excited states and assuming that these
excited states are not thermally accessible, the shielding tensor \mat{\sigma}
can be divided in two parts: $\mat{\sigma}=\mat{\sigma}^\text{p}+
\mat{\sigma}^\text{r}$, the first part representing the `paramagnetic shift',
which is due entirely to the (quasi-)degenerate ground state and which can be
calculated from knowledge of the ground-state wave functions only; the second
part representing the `Ramsey term', which is the only term in case of a
non-degenerate ground state.  These terms are distinguished by the fact that in
the limit of vanishing zero-field splitting of the ground state
$\mat{\sigma}^\text{r}$ is temperature independent while
$\mat{\sigma}^\text{p}$ is proportional to $1/T$. The present paper will
consider the effect of a zero-field splitting on $\mat{\sigma}^\text{p}$.

The electronic Hamiltonian consists of two parts: $H=H_0+V$. Here $V$ is the
perturbation due to the applied magnetic field and the magnetic field arising
from the magnetic moments of the nuclei. $H_0$ is the Hamiltonian in the
absence of these fields, but including all other relevant electronic
interactions. This means that $H_0$ includes those interactions that are
responsible for the zero-field splitting of the ground state. If the ground
manifold consists of $\omega$ states \ket{\psi_{\lambda a}}, eigenstates of
$H_0$ with energies $E_\lambda$, an expression of $H_0$ valid within this
manifold is:
\begin{equation*}
H_0=\sum_{\lambda,a}^\omega E_\lambda\,\ket{\psi_{\lambda a}}\bra{\psi_{\lambda a}}
\end{equation*}
Here the index $\lambda$ counts the energy levels of the manifold, and the index
$a$ labels a basis in case $E_\lambda$ is degenerate. For our present purpose,
the term $V$ consists of two perturbations that combined give rise to
paramagnetic shielding: the electronic Zeeman interaction
$V_\text{z}=-\vec{m}\cdot\vec{B}$, and the hyperfine coupling
$V_\text{hf}=\bm{\mathcal{F}}\cdot\bm{\mu}$, where \vec{m} is the electronic
magnetic moment, \vec{B} is the applied field, $\bm{\mathcal{F}}$ is the
hyperfine field and $\bm{\mu}$ is the nuclear magnetic moment
\cite{Abragam1961}.

A general formula for the shielding tensor was proposed in Ref.\
\onlinecite{VandenHeuvel2012a}, to which the reader is referred for more details:
\begin{equation}\label{sigma-def}
\sigma_{ij}=\left.\frac{\partial^2F}{\partial B_i\partial \mu_j}\right\vert_0
\end{equation}
Here $F$ is the electronic Helmholtz free energy of the full Hamiltonian
$H=H_0+V$. Evaluation of \eqnref{sigma-def},
and retention of the temperature-dependent paramagnetic part only, 
 leads to\cite{VandenHeuvel2012a}:
\begin{equation}\label{sigma1}
\sigma^\text{p}_{ij}= \left\langle\int_0^\beta
  e^{wH_0}m_ie^{-wH_0}\mathcal{F}_jdw\right\rangle_0,
\end{equation}
where $\langle\cdot\rangle_0$ stands for the thermal average in the canonical
ensemble of $H_0$, and $\beta=1/{kT}$.
The expression \eqnref{sigma1} can now be easily integrated\cite{VandenHeuvel2012a} leading 
to an exact sum-over-states formula:
\begin{multline}\label{sigmap-gen}
\sigma^\text{p}_{ij}=\frac{1}{Q_0} \sum_\lambda e^{-\beta E_\lambda} 
\biggl[\beta\sum_{a,a'}\braket{\psi_{\lambda a}}{m_i}{\psi_{\lambda a'}} 
  \braket{\psi_{\lambda a'}}{\mathcal{F}_j}{\psi_{\lambda a}}\\
  +2\sum_{\lambda'\neq \lambda}\sum_{a,a'} \frac{\braket{\psi_{\lambda a}}
  {m_i}{\psi_{\lambda' a'}}\braket{\psi_{\lambda' a'}}{\mathcal{F}_j}
  {\psi_{\lambda a}}}{E_{\lambda'}-E_\lambda}\biggr].
\end{multline}
Here $Q_0=\sum_{\lambda,a}e^{-\beta E_\lambda}$ is the
partition function.

%\section{Application to a spin state with zero-field splitting}

Our aim here is to rewrite \eqnref{sigmap-gen} for the case of a pure spin ground multiplet 
split as a result of spin-orbit coupling. In molecules composed of light atoms the spin-orbit 
coupling is often a small perturbation, whose effect can be treated to good accuracy in the lowest order
of perturbation theory.  It is then a well known result \cite{Abragam_EPR} that
such treatment leads to a spin Hamiltonian 
\begin{equation*}\label{spinham}
H= \vec{S}\cdot\mat{D}\cdot\vec{S}+\bohr \vec{B}\cdot\mat{g}\cdot\vec{S} +
\vec{S}\cdot\mat{A}\cdot\vec{I}.
\end{equation*}
In the notation introduced in the previous section, we thus have
\begin{equation*}
H_0=\vec{S}\cdot\mat{D}\cdot\vec{S},\qquad m_i= -\bohr\sum_j{g_{ij}}S_j, \qquad
\mathcal{F}_i= \frac{1}{g_I \nbohr}\sum_j A_{ji} S_j,
\end{equation*}
where the indices label the Cartesian directions $x,y,z$ and we have used
$\bm{\mu}=g_I\nbohr\vec{I}$ to convert between the nuclear magnetic moment and
the nuclear spin. 

The states to be used in \eqnref{sigmap-gen} can then be easily found by diagonalizing 
$H_0$ in the space of $2S+1$ spin basis states \ket{S\,M}. Naturally the eigenfunctions 
will depend on the zero-field splitting tensor \mat{D}. For the moment we leave the latter
unspecified and denote the eigenfunctions generically by \ket{S\,\lambda a}.
Now applying \eqnref{sigmap-gen} gives
\begin{multline}\label{sigmap-spin}
\sigma^\text{p}_{ij}=-\frac{\bohr}{g_I \nbohr}\frac{1}{Q_0} \sum_{kl}
g_{ik}A_{lj} \sum_\lambda e^{-\beta E_\lambda}
\biggl[\beta\sum_{a,a'}\braket{S\,\lambda a}{S_k}{S\,\lambda a'}
\braket{S\,\lambda a'}{S_l}{S\,\lambda a}\\
+2\sum_{\lambda'\neq \lambda}\sum_{a,a'} \frac{\braket{S\,\lambda a}{S_k}{S\,\lambda' a'}
\braket{S\,\lambda' a'}{S_l}{S\,\lambda a}}{E_{\lambda'}-E_\lambda}\biggr].
\end{multline}
\eqnref{sigmap-spin} represents the main result of this communication. 

\eqnref{sigmap-spin} has to be compared with the formula proposed by Pennanen
and Vaara in Ref.\ \onlinecite{Pennanen2008}. Their Eq.\ (10) reads
\begin{equation}\label{vaara1}
\sigma^\text{p}_{ij}=-\frac{\bohr}{g_I \nbohr}\frac{1}{kT}\sum_{kl}g_{ik}A_{lj} \langle
S_kS_l\rangle_0,
\end{equation}
which can be written more explicitly by performing the thermal average over the eigenfunctions of $H_0$:
\begin{equation}\label{vaara2}
\begin{split}
\sigma^\text{p}_{ij}&=-\frac{\bohr}{g_I \nbohr}\frac{\beta}{Q_0}
\sum_{kl}g_{ik}A_{lj}\sum_\lambda e^{-\beta E_\lambda} \sum_a\braket{S\,\lambda
a}{S_kS_l}{S\,\lambda a}\\
 &= -\frac{\bohr}{g_I \nbohr}\frac{\beta}{Q_0} \sum_{kl}g_{ik}A_{lj}\sum_\lambda 
 e^{-\beta E_\lambda} \sum_{\lambda',a,a'}\braket{S\,\lambda a}{S_k}{S\,\lambda' a'}
 \braket{S\,\lambda' a'}{S_l}{S\,\lambda a}.
\end{split}
\end{equation}
Clearly, Eqs.\ \eqref{sigmap-spin} and \eqref{vaara2} are not equal. In fact,
only in the special case $\mat{D}=0$, i.e.\ in the absence of zero-field splitting, the
two expressions Eqs.\ \eqref{sigmap-spin} and \eqref{vaara2} lead to the 
same formula for the paramagnetic shielding tensor:
\begin{equation*}
\bm{\sigma}^\text{p}=-\frac{\bohr}{g_I \nbohr}\frac{1}{kT}
\frac{S(S+1)}{3}\mat{g}\mat{A}.
\end{equation*}
In every other case we argue that the correct formula is given by \eqnref{sigmap-spin}.
Note that $\sigma^\text{p}_{ij}$ can be expressed in a form
that is only similar to \eqnref{vaara1}, if we take the thermal average
of a different operator:
\begin{equation}\label{integral}
\sigma^\text{p}_{ij}=-\frac{\bohr}{g_I \nbohr}\frac{1}{kT}\sum_{kl}g_{ik}A_{lj}
\left\langle \int_0^\beta e^{\tau H_0}S_ke^{-\tau H_0}S_l
\,d\tau\right\rangle_0.
\end{equation}

%\subsection{Example: triplet state}

Finally, to illustrate the difference between the results presented in this communication
and previous works, and to argue for the correctness of our proposal, we consider the simple 
but very common case of a triplet state with an axial zero-field splitting $H_0=DS_z^2$, 
and axial $g$- and $A$-tensors:
\begin{equation*}
\mat{g}=\begin{pmatrix} g_\bot&0&0\\
                        0&g_\bot&0\\
                        0&0&g_{||}
        \end{pmatrix},\qquad
\mat{A}=\begin{pmatrix} A_\bot&0&0\\
                        0&A_\bot&0\\
                        0&0&A_{||}
        \end{pmatrix}.
\end{equation*}
The eigenstates of $H_0$ are simply the \ket{S\,M} (with $S=1$), which we
further denote by their $M$ value alone. Thus we have two energy levels: \ket{0} at
energy 0, and \ket{\pm1} at energy $D$. On evaluating the newly proposed \eqnref{sigmap-spin} we
find:
\begin{equation}\label{triplet}
\begin{split}
\sigma^\text{p}_\bot&=-\frac{\bohr}{g_I \nbohr} \frac{2g_\bot A_\bot}{D}
\frac{1-e^{-\beta D}}{1+2e^{-\beta D}}\\
\sigma^\text{p}_{||}&=-\frac{\bohr}{g_I \nbohr}\frac{2g_{||} A_{||}}{kT}
\frac{e^{-\beta D}}{1+2e^{-\beta D}}.
\end{split}
\end{equation}
Previously proposed \eqnref{vaara1} on the other hand predicts:
\begin{equation}\label{triplet-vaara}
\begin{split}
\sigma^\text{p}_\bot&=-\frac{\bohr}{g_I \nbohr} \frac{g_\bot A_\bot}{kT}
\frac{1+e^{-\beta D}}{1+2e^{-\beta D}}\\
\sigma^\text{p}_{||}&=-\frac{\bohr}{g_I \nbohr}\frac{2g_{||} A_{||}}{kT}
\frac{e^{-\beta D}}{1+2e^{-\beta D}},
\end{split}
\end{equation}
which disagrees with \eqnref{triplet} on the value of $\sigma^\text{p}_\bot$.
Note that $\sigma^\text{p}_{||}$ is the same in both theories only because of the
specific axial symmetry of this system, implying that the ZFS Hamiltonian commutes with 
the component of the spin operator along the axial direction.
That the formula for $\sigma^\text{p}_\bot$ in \eqnref{triplet-vaara} must be
wrong can be deduced by considering the low-temperature limit $kT\ll D$ for
$D>0$ (easy-plane ZFS anisotropy).  In this situation the ground state is \ket{0} and is the only populated
state of the system. Therefore the shielding should be temperature-independent.
\eqnref{triplet-vaara} however, predicts a Curie behaviour in this limit: 
\[ \sigma^\text{p}_\bot\rightarrow -\frac{\bohr}{g_I \nbohr} \frac{g_\bot
A_\bot}{kT}. \] The correct limit is obtained from \eqnref{triplet} and is
indeed a constant: 
\[\sigma^\text{p}_\bot\rightarrow -\frac{\bohr}{g_I \nbohr} \frac{2g_\bot
A_\bot}{D}. \]

%\section*{Acknowledgments}
A.S. acknowledges support from the Early Career Researcher Grant (ECR 2012) from the 
University of Melbourne.

%\section*{References}

\end{document}